\documentclass[aps,prl,final,notitlepage,sort&compress,twocolumn]{elsarticle}
\usepackage{amsmath}
\usepackage{amssymb}
\usepackage{graphicx}
\usepackage{subfigure}
\usepackage{appendix}
\usepackage{txfonts}
\usepackage{lineno}
\begin{document}

 \newcommand{\re}{\mathop{\mathrm{Re}}}
 \newcommand{\im}{\mathop{\mathrm{Im}}}
 \newcommand{\D}{\mathop{\mathrm{d}}}
 \newcommand{\I}{\mathop{\mathrm{i}}}
 \newcommand{\E}{\mathop{\mathrm{e}}}
 \newcommand{\unite}[2]{\mbox{$#1\,{\rm #2}$}}
 \newcommand{\myvec}[1]{\mbox{$\overrightarrow{#1}$}}
 \newcommand{\mynor}[1]{\mbox{$\widehat{#1}$}}
 \newcommand{\rmsemit}{\mbox{$\tilde{\varepsilon}$}}
 \newcommand{\meanavg}[1]{\mbox{$\langle{#1}\rangle$}}

\title{Beam dynamics performances and applications of a low-energy \\
electron-beam magnetic bunch compressor}

\author[niu]{C. R. Prokop}
\author[niu,apc]{P. Piot}
\author[lanl]{B. E. Carlsten}
\author[ad]{M. Church}
\address[niu]{Northern Illinois Center for Accelerator \& Detector Development and Department of Physics, \\
Northern Illinois University, DeKalb, IL  60115, USA}
\address[apc]{Accelerator Physics Center, Fermi National Accelerator Laboratory, Batavia, IL  60510, USA}
\address[lanl]{Acceleration Operations and Technology Division, Los Alamos National Laboratory, Los Alamos, NM 87545, USA}
\address[ad]{Accelerator Division, Fermi National Accelerator Laboratory, Batavia, IL  60510, USA}
\date{today}

\begin{abstract}
Many front-end applications of electron linear accelerators rely on the production of temporally-compressed bunches. The shortening of electron bunches is often realized with magnetic bunch compressors located in high-energy sections of accelerators. Magnetic compression is subject to collective effects including space charge and self interaction via coherent synchrotron radiation. In this paper we explore the application of magnetic compression to low-energy ($\sim 40$~MeV), high-charge (nC) electron bunches with low normalized transverse emittances ($< 5$~$\mu$m). 
\end{abstract}


\maketitle

\section{Introduction}
Most of the photoinjectors being used for generation of bright electron bunches for, e.g., free-electron laser (FEL) applications consist of generating and rapidly accelerating the electron bunch to high energy and subsequently shortening the bunch using a magnetic bunch compressor~\cite{carlsten}. The only deviation to such a design is the combined acceleration and compression using velocity bunching~\cite{ferrario}. Attempts to operate low-energy magnetic bunch compressors have to date been inconclusive~\cite{braun} or deemed incompatible with the production of low-emittance beams~\cite{BCTTF1}. In this paper we explore and demonstrate via numerical simulations that low-energy bunch compression performed on a $\sim 40$-MeV electron beam can be viable depending on requirements. We especially present trade-off curves between transverse emittance and peak current for several cases of electron-bunch charge. In addition, our simulations are performed with several computer programs thereby enabling a benchmarking of very different approaches for modeling collective effects and especially coherent synchrotron radiation (CSR)~\cite{vladimir,rui}. Our study considers the magnetic bunch compression planned in the 40-50~MeV photoinjector of the Advanced Superconducting Test Accelerator (ASTA) currently under construction at Fermilab~\cite{Nagaitsev,leibnitzIPAC12}. 

\section{Accelerator beamline overview}
\begin{figure}[hb!!!!!!!!!!!]
\centering
\includegraphics[width=0.50\textwidth]{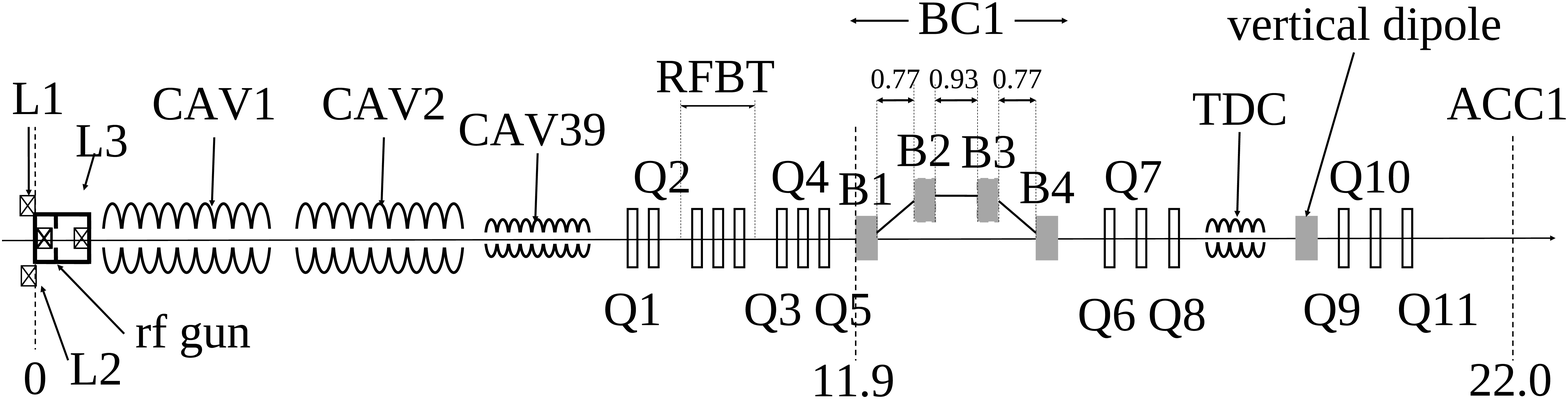}
\caption{\label{fig:InjectorOverview} Injector configuration at ASTA.  The  ``RF gun", ``L1" and ``L2" respectively correspond to the gun cavity and surrounding solenoid magnets, ``CAV1",  ``CAV2", and ``CAV39" are superconducting RF cavities, ``RFBT" is the round-to-flat beam transformer, and ``BC1" refers to the magnetic bunch compressor, and B1-4 are the dipoles of the chicane, with distance between the dipoles marked in the figure. The number below the beamline indicates the axial positions in meters w.r.t. the photocathode surface.}
\end{figure}

The compression of a $\sim 40$-MeV electron bunch via magnetic compression is investigated for the case of the ASTA photoinjector diagrammed in Fig.~\ref{fig:InjectorOverview}. The beamline includes a  photoemission electron source consisting of a cesium telluride (Cs$_2$Te) photocathode located on the back plate of a 1+1/2 cell radiofrequency (RF) cavity operating at 1.3~GHz~\cite{dwersteg}. The cathode is illuminated with a 3-ps ultraviolet laser pulse with uniform radial distribution and a Gaussian temporal profile. The RF gun is surrounded by two solenoidal lenses that control the beam's transverse size and emittance. Downstream of the RF gun, the typical beam energy is $\sim 5$~MeV. The bunches are further accelerated up to 50 MeV by two 1.3-GHz superconducting RF (SCRF) accelerating cavities (labeled as CAV1 and CAV2 in Fig.~\ref{fig:InjectorOverview}). A third  SCRF cavity (CAV39) operating at 3.9 GHz will eventually be incorporated to correct for nonlinear longitudinal phase space distortions~\cite{smith,dowell,piot3rd}. Because of its superconducting nature, the ASTA facility produces electron bunches repeated at 3 MHz arranged in a 1-ms 5-Hz RF macropulse. The downstream beamline includes quadrupoles, steering dipole magnets, and diagnostics stations. A skew-quadrupole channel can be set up as a round-to-flat-beam transformer (RFBT) to convert an incoming angular-momentum-dominated beam into a flat beam with high transverse emittance ratio~\cite{yine2004,flat2006}. The beamline also incorporates a four-bend magnetic bunch compressor (BC1) which, consists of four 0.2-m rectangular dipoles (B1, B2, B3, B4) with respective bending angles of (+,-,-,+) $18^{\circ}$. The longitudinal dispersion of BC1 is $R_{56}=-0.19$~m. Finally a single-shot longitudinal phase space diagnostics combining a transverse-deflecting cavity (TDC) with a vertical spectrometer will be installed~\cite{prokopLPS}.  

\begin{figure}[h]
\centering
\includegraphics[width=0.48\textwidth]{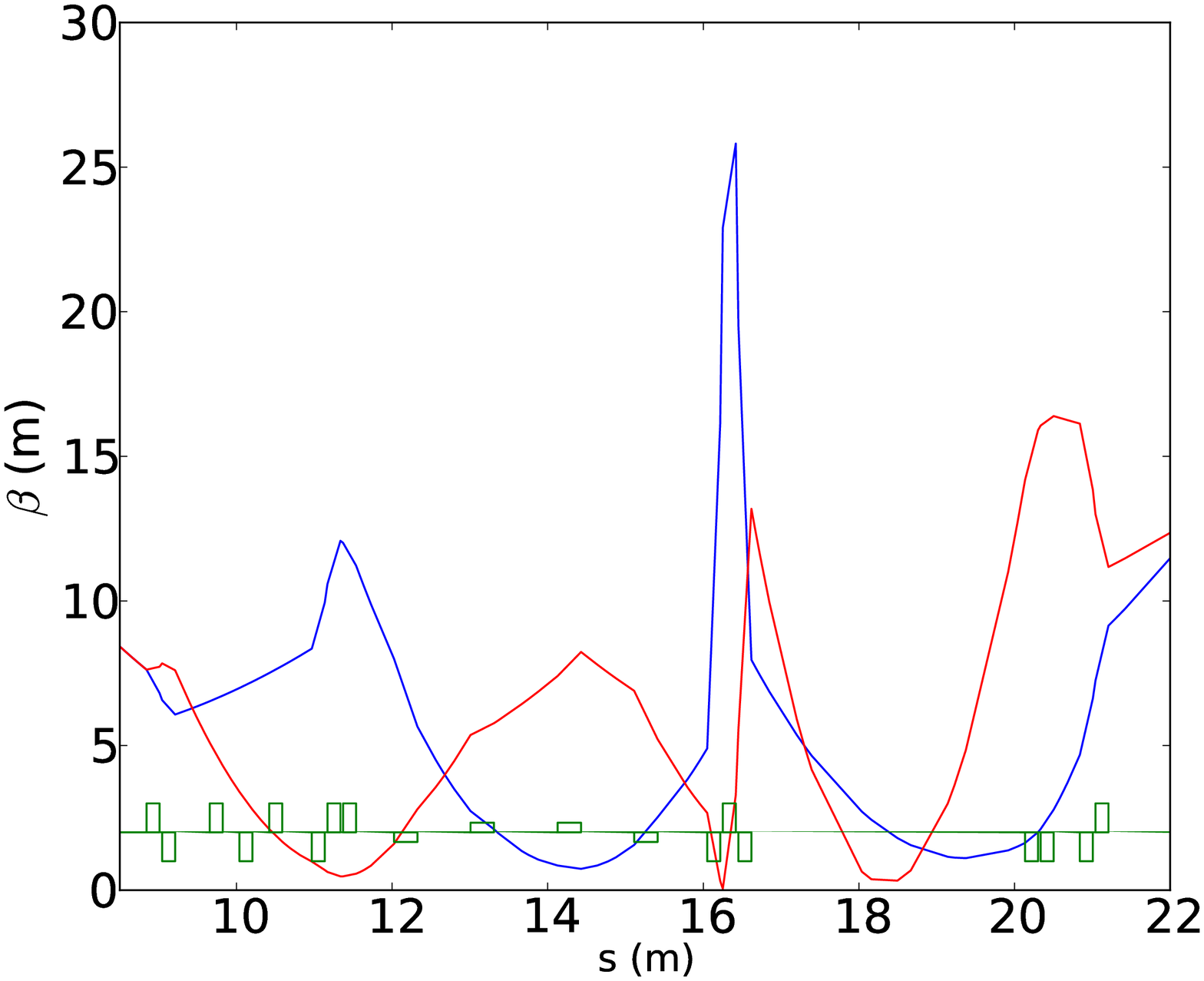}
\caption{\label{fig:latticefunc} Evolution of the horizontal (blue) and vertical (red) betatron functions through the ASTA injector. The green rectangles indicate the location of quadrupole and dipole (smaller rectangles) magnets. The BC1 compressor is located at $s\in [11.9, 15.1]$~m. The origin of the horizontal axis ($s=0$~m, not shown) corresponds to the photocathode surface. (color online)}
\end{figure}

The beam dynamics through CAV2 were simulated with {\sc astra} and optimized using a genetic optimizer for several cases of charge and photocathode drive-laser configurations; see Ref.~\cite{piotIPAC10}.  The resulting phase space distributions are used as a starting point for transport and compression through the beamline downstream of CAV39. The quadrupoles settings were optimized for the various operating charges using the single-particle dynamics program {\sc elegant}~\cite{elegant}. The evolution of the nominal betatron functions downstream of CAV39 up to the cryomodule entrance is plotted in Fig.~\ref{fig:latticefunc}. 

\section{Modeling methodology}

%

The bunch-compression performance of BC1 was explored via numerical simulations using {\sc impact-z}~\cite{ImpactZ} and {\sc csrtrack}~\cite{CSRtrack}.  Both programs model the beam as an ensemble of interacting macroparticles and integrate the equations of motion to advance the macroparticles along a user-specified beamline.   

In {\sc impact-z} the space-charge (SC) interaction is modeled using  a mean-field quasi-static particle-in-cell (PIC) algorithm and the (point-like) macroparticles are advanced through the beamline using high-order transfer maps. Each beamline element is segmented into axial slices modeled by transfer maps. Between each transfer-map segment, {\sc impact-Z} applies a space-charge ``kick" evaluated from the mean-field PIC SC algorithm~\cite{ImpactZ}. CSR effects are included in {\sc impact-Z} using the one-dimensional formulation described in Ref.~\cite{SaldinBig1997}.  The one-dimensional model is valid provided~\cite{DerbenevCriteria}
\begin{eqnarray}
\label{eq:DerbenevCriterion}
{\cal D} (s) \ll 1 \mbox{, with~} {\cal D}(s) \equiv \frac{\sigma_x(s)}{\sigma_z(s)} \sqrt{\frac{\sigma_x(s)}{R(s)}}, 
\end{eqnarray}
where $R(s)$ is the trajectory's radius of curvature and  $\sigma_x(s)$ and $\sigma_z(s)$ are respectively the transverse  and longitudinal root-mean-square (RMS) sizes at the curvilinear beamline position $s$.

In {\sc impact-z} the longitudinal charge distribution needed for the 1D CSR model is obtained from a longitudinal binning of the macroparticle ensemble. Convergence studies were carried out in order to determine the optimal number of longitudinal bins, $N_z$, to be used for both the SC and CSR calculations.  Low values of $N_z$ generally underestimated the peak-current and therefore the collective effects, while large values of $N_z$ introduce numerical noise that can lead to artifacts (e.g. numerically-induced microbunching)~\cite{mubunching}.  The convergence study~\cite{ProkopBC1TM} revealed an appropriate value of $N_z= 256$ for a bunch represented by $N=2 \times 10^5$ macroparticles. The number of bins in the transverse dimensions was set to $N_x=N_y=16$. The {\sc impact-z} simulations presented in the rest of this paper use this set of parameters. 

The program {\sc csrtrack} was specifically developed to efficiently simulate the impact of bunch radiative self-interaction via CSR.  {\sc Csrtrack} incorporates several models including a  2D particle-to-particle (P2P) model that directly computes the forces on macroparticles from the Li\'enard-Wiechart potentials evaluated at retarded times. These calculations are self-consistent and enable the computation of both the transverse and longitudinal force contributions from SC and CSR effects. Since the P2P model is computationally intensive, (the calculation time scales as $N^2$), {\sc csrtrack} also includes an improved one-dimensional model referred to as the 1D Projected (1DP) model. The 1DP model uses the 1D projection of the smoothed charge distribution convoluted with a kernel function \cite{DohlusCSRMethods}. Compared to the model of Ref.~\cite{SaldinBig1997}, the 1DP model is not limited to the ultra-relativistic regime. {\sc Csrtrack}'s P2P model treats each macroparticle as a 3D Gaussian charge distribution (referred to as ``sub-bunch'')  in the $(x,y,z)$ space with distribution 
\begin{eqnarray}
g(x,y,z)=\frac{1}{(2\pi)^{3/2}\sigma_{h}\sigma_{v}\sigma_{||}}e^{-\frac{x^2}{2\sigma_{h}^2}-\frac{y^2}{2\sigma_{v}^2}-\frac{z^2}{2\sigma_{||}^2}},
\end{eqnarray}
where $\sigma_h$, $\sigma_v$ and $\sigma_{||}$ are respectively the horizontal, vertical and longitudinal RMS sizes of the sub-bunches.    The resulting beam's spatial charge distribution is $\Phi(x,y,z)=\sum_{j=1}^N Q_j g(x-x_j, y-y_j,z-z_j)$ where $Q_j$ and $(x_j, y_j, z_j)$ are respectively the $i$-th sub-bunch charge and LPS coordinates. In {\sc csrtrack}, $\sigma_v$ and $\sigma_{||}$ may be defined relative to the vertical and longitudinal RMS bunch sizes, $\sigma_y$ and $\sigma_z$, respectively, and are adjusted along the bunch compressor as the dimensions of the bunch change.  Due to the computational intensiveness of the P2P model,  only $10^4$ sub-bunches were used  compared to $2\times10^5$ used with the 1DP model. This relatively-low number of sub-bunches requires a large $\sigma_{||}$ of 5$\%$ of the longitudinal bunch length $\sigma_z$ for the 1DP simulations and 10$\%$ for the P2P simulations; for a detailed study see Ref.~\cite{ProkopBC1TM}.  The P2P model requires parameters for the sub-bunches in the horizontal and vertical dimensions, $\sigma_{h}$ and $\sigma_{v}$.  We chose $\sigma_{v}$=~0.10$\sigma_y$ and  $\sigma_{||}$=~0.10$\sigma_z$.  However, {\sc csrtrack} does not allow $\sigma_{h}$ to be set as a variable of the horizontal RMS width, so we instead chose 0.1~mm, which is on the order of 10\% of the RMS size $\sigma_x$ for all four of the bunch charges presented here.  Both of {\sc csrtrack}'s models neglect collective forces in the vertical dimension.

The {\sc astra} simulations of the beam generation and acceleration simulated up to $s=8.2$~m from the photocathode were used as a starting point for our simulations. A {\sc python} program, {\sc gluetrack}~\cite{igor}, was used to manipulate the beam distributions and generate macroparticle distributions suitable for {\sc impact-z} and {\sc csrtrack}. For all the bunch-compressor studies, {\sc impact-z} was used to track the bunch distribution from $s=8.2$~up to $s=11.8$~m corresponding to 0.1~m upstream of the entrance face of the first dipole (B1) of BC1. This distribution was used as a starting point for the bunch-compressor simulations, which include BC1 and a 1.0-m downstream drift to allow for the influence of SC possible transient CSR effects. 

Simulations were performed for four  cases of bunch charges ranging from  3.2~nC to 20~pC. For each charge the transverse emittance was optimized with {\sc astra}~\cite{piotIPAC10}.  The distributions were manipulated using {\sc gluetrack} to adjust several parameters including their Courant-Snyder (C-S) parameters, and longitudinal phase space (LPS) chirp ${\cal C} \equiv -\meanavg{z_i\delta_i}/\sigma_{z,i}^2$ where $(z_i,\delta_i)$ are the coordinates in the LPS, the $\meanavg{u}$ indicates the statistical averaging of variable $u$ over the LPS distribution and $\sigma_{z,i}\equiv \meanavg{z_i^2}^{1/2}$. We also modeled the effect of CAV39 by numerically removing the second order correlation in the LPS distribution. 

\begin{table}[h!]
\caption{\label{tab:OpticsParameters} Transverse and longitudinal beam parameters 0.1-m upstream of B1 dipole entrance face. Only the Courant-Snyder parameters were fixed while the other parameters depend on the bunch charge or upstream beamline settings.}
\begin{center}
\begin{tabular}{c c c}
\hline 
Parameter	 & Value & Units\\
\hline \hline
$\beta_{x,i} $	& 8	& m  \\
$\alpha_{x,i}$	& 3	& - \\
$\beta_{y,i} $  & 1.6	& m \\
$\alpha_{y,i}$	& -1.6	& - \\
${\cal C}$	& [1.0,6.0] & m$^{-1}$ \\
total energy 	& 38.6	& MeV \\
\hline 
\end{tabular}
\end{center}
\end{table}

\begin{figure}[h!]
\centering
\includegraphics[width=0.47\textwidth]{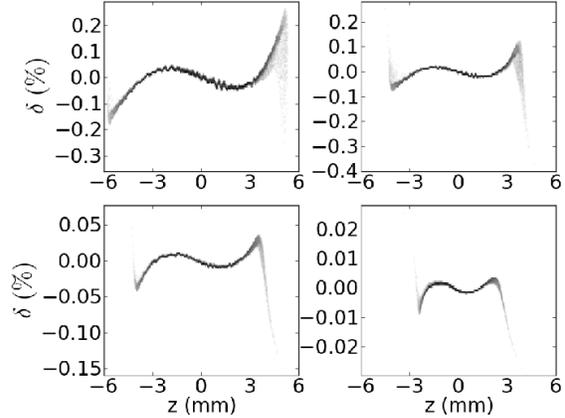}
\caption{LPS distributions 0.1-m upstream of dipole magnet B1's entrance face for 3.2 (a), 1.0 (b), 0.25 (c) and 0.02~nC (d) bunches. The distributions were obtained from simulations of the photoinjector beam dynamics in {\sc astra}; see Ref.~\cite{piotIPAC10}. The ordinates $z>0$ correspond to the head of the bunch.}\label{fig:InitialLPS}
\end{figure}

The charge-independent beam parameters computed 0.1-m upstream of dipole magnet B1 are summarized in Tab.~\ref{tab:OpticsParameters}. The beam distribution was matched to achieve the transverse C-S parameters listed in Tab.~\ref{tab:OpticsParameters} (see discussion below) upstream of BC1. The initial LPS chirp was tuned by removing the second-order correlation and scaling the first-order correlation between $\delta_i$ and $z_i$. The other LPS parameters and emittances (shown for the four charges in Tab.~\ref{tab:ChargeParameters}) are inherent to the generation process and were not adjusted.  The initial LPS distribution for each of the four charges appears in Fig.~\ref{fig:InitialLPS} with its linear correlation removed.  The $S$-shaped LPS is a remnant of space charge effects during the bunch generation and transport before acceleration in CAV1 and CAV2~\cite{loulergue}. As expected larger charges yield higher total fractional momentum spread.

\begin{table}[h!]
\caption{\label{tab:ChargeParameters} Initial normalized  transverse $\varepsilon_{x/y,i}$ and longitudinal $\varepsilon_{z,i}$ emittances and RMS bunch length $ \sigma_{z,i}$ for the four cases of charge considered in this paper. The parameters are computed 0.1-m upstream of dipole magnet B1's  entrance face.}
\begin{center}
\begin{tabular}{c c c c c}
\hline 
Q (nC) & $\varepsilon_{x,i}$ ($\mu$m) & $\varepsilon_{y,i}$ ($\mu$m)  &  $\varepsilon_{z,i}$ ($\mu$m)  & $ \sigma_{z,i}$ (mm) \\
\hline \hline 
3.2 & 4.43 & 4.58 & 82.19 & 2.56\\
1.0 & 2.20 & 2.22 & 33.41 & 1.95\\
0.250 & 0.580 & 0.576 & 14.37 & 1.93\\
0.020 & 0.296 & 0.297 & 2.54 & 1.26 \\
\hline 
\end{tabular}
\end{center}
\end{table}

Simulations were performed for LPS chirps ${\cal C} \in [1.0,6.0]$~m$^{-1}$ with the bulk of the simulations performed around the maximum-compression value ${\cal C}=-1/R_{56} \simeq 5.2$~m$^{-1}$, corresponding to enhanced collective effects.

\begin{figure}[h!]
\centering
\includegraphics[width=0.55\textwidth]{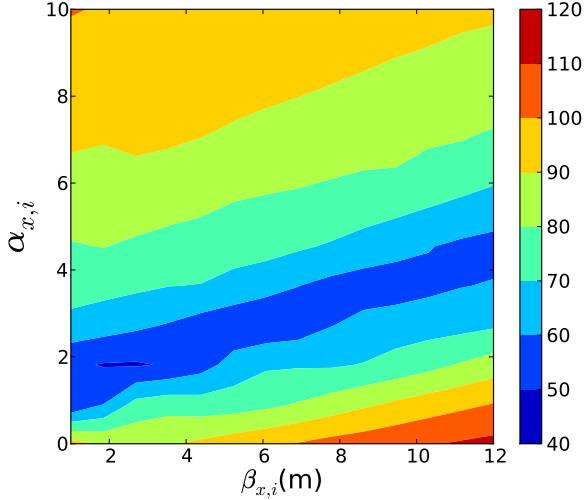}
\caption{Contour plot of the final normalized horizontal emittance ($\varepsilon_{x}$ in $\mu$m) as a function of the C-S parameters $\beta_{x,i}$ and $\alpha_{x,i}$ 0.1-m upstream of dipole B1. The simulations were performed with {\sc csrtrack}'s 1D-Projected model and a bunch charge of 3.2~nC . From this data, we selected the values for our simulations, $(\beta_{x,i},\alpha_{x,i})=(8.0~\mbox{m}, 3.0)$.   (color online)}.\label{fig:TwissScan}
\end{figure}

\begin{figure}[h!]
\centering
\includegraphics[width=0.48\textwidth]{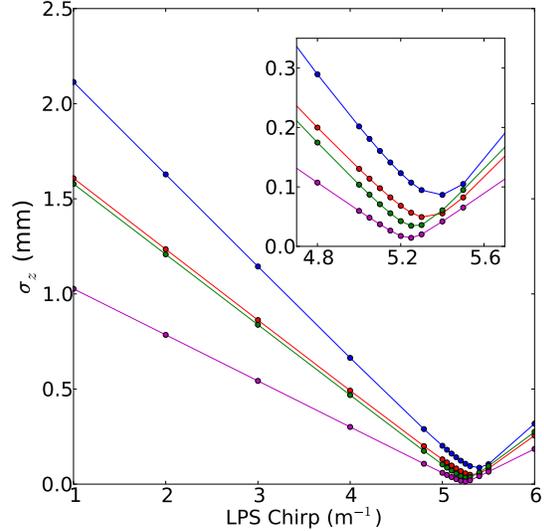}
\caption{RMS bunch length $\sigma_z$ downstream of BC1 as a function of the LPS chirp for various bunch charges using {\sc Impact-Z}'s CSR+SC model model, for 3.2-nC (blue), 1.0-nC (red),  250-pC (green),  20-pC (magenta) bunch charges. The inset plot corresponds to a close-up around chirp values that achieve minimum RMS bunch lengths.  (color online)}\label{fig:IZ_sigz}
\end{figure}

The initial values for the betatron functions were selected such that the beam experiences a waist between the third and fourth dipole~\cite{martin05}.  Simulations performed with the 1DP {\sc csrtrack} model also confirmed there is a set of horizontal C-S parameters that minimizes the bending-plane emittance growth as displayed in Fig.~\ref{fig:TwissScan}.  The upstream magnets were tuned to provide the incoming horizontal C-S parameter $(\beta_{x,i},\alpha_{x,i})=(8.0~\mbox{m},3.0)$ shown in Fig.~\ref{fig:latticefunc}.  Lastly, the final RMS bunch lengths as a function of the initial LPS chirp are given in Fig.~\ref{fig:IZ_sigz} as obtained using {\sc impact-z}'s SC+CSR model.  In addition, there are slight variations in the  bunch length between the four different models due to their collective effects inhibiting compression to varying degrees.

\section{Benchmarking of numerical models}

The beam dynamics simulations throughout BC1 were performed for several degrees of bunch compression (controlled with the LPS chirp) for the four cases of bunch charges that will be used at the ASTA. Four simulation algorithms were used as described earlier.  Because of its computational effort, {\sc crtrack}'s P2P model  simulations were restrained to a smaller range of values for ${\cal C}$.

\begin{figure}[h!]
\centering
\includegraphics[width=0.48\textwidth]{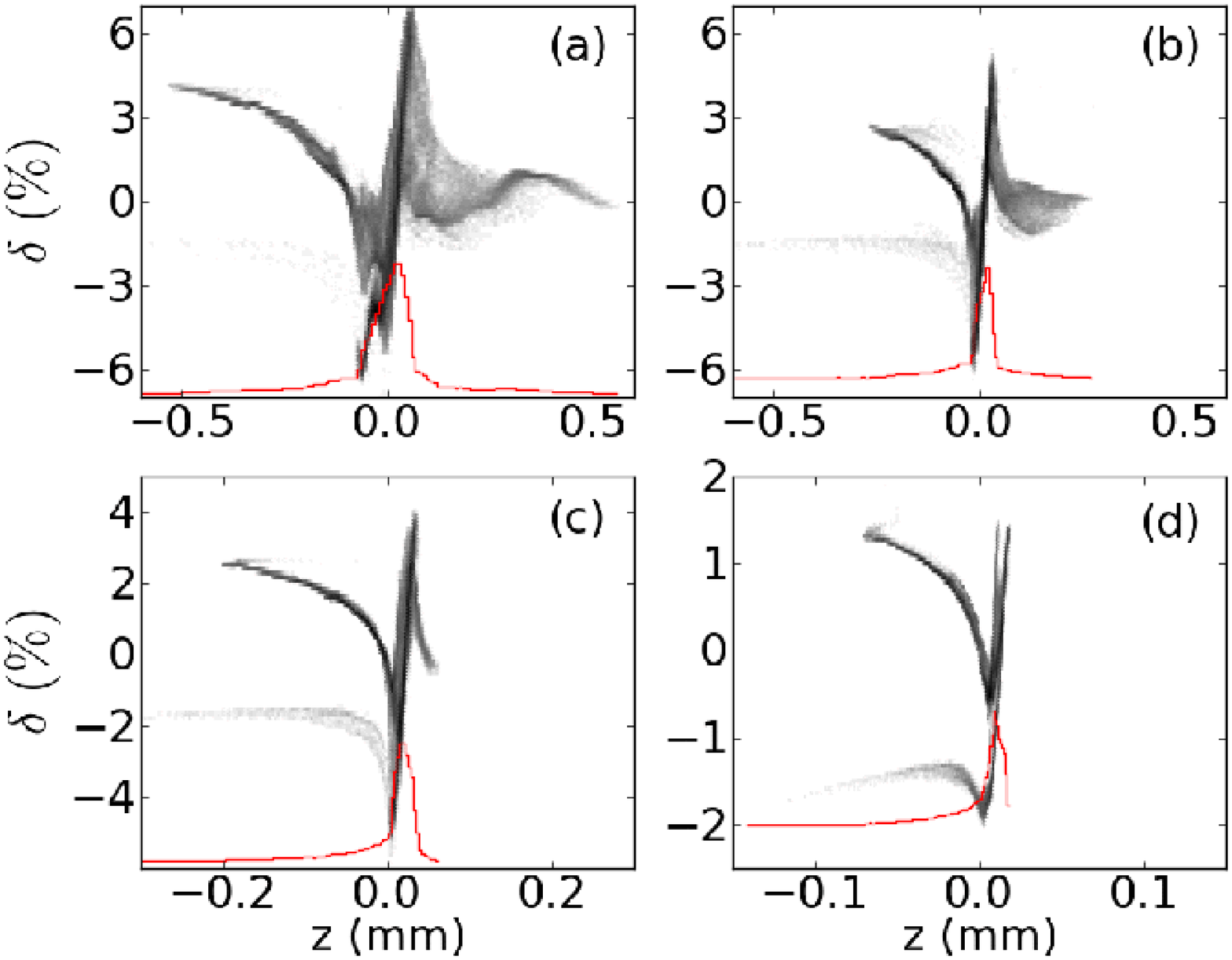}\\
\includegraphics[width=0.48\textwidth]{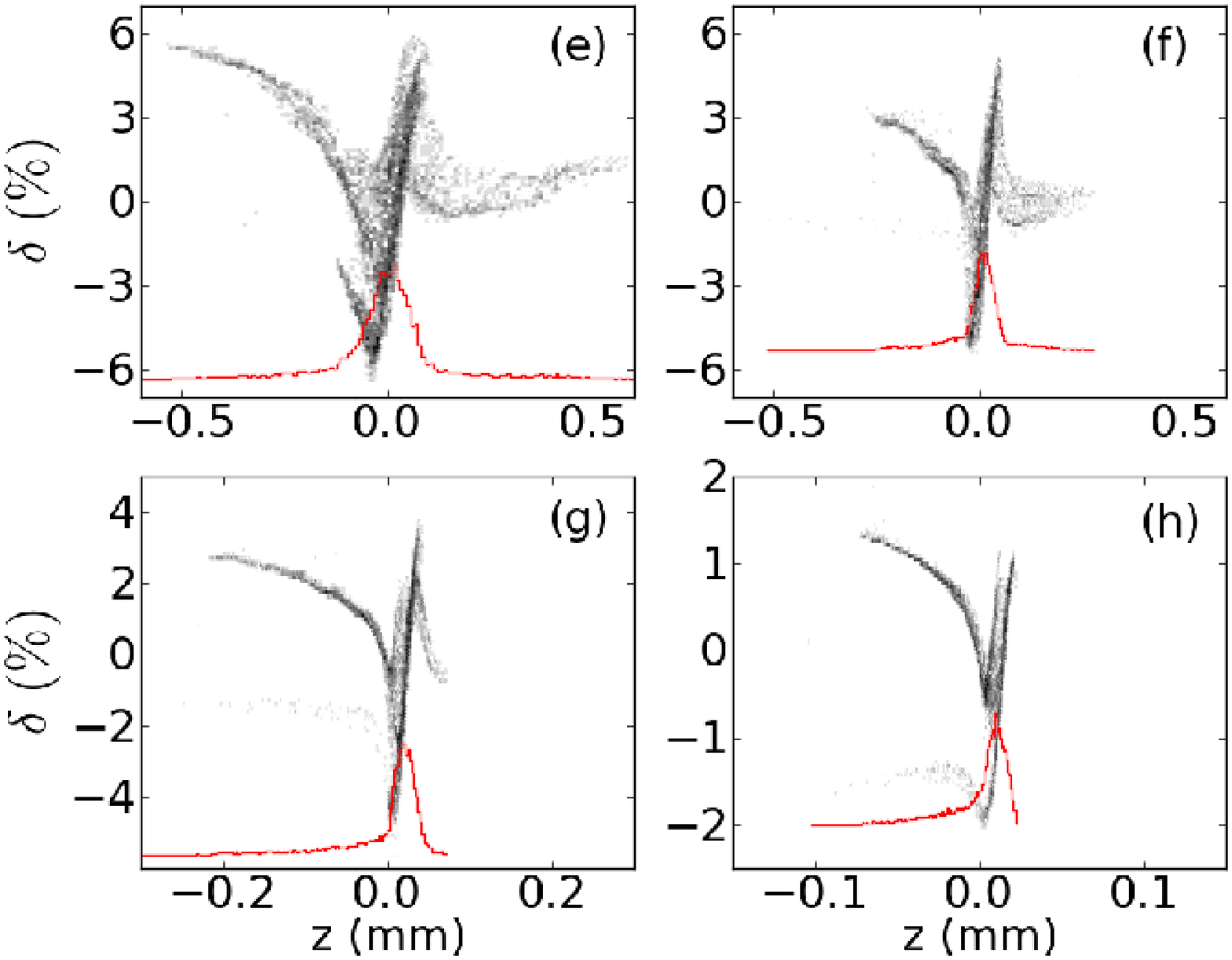}
\caption{\label{fig:LPScompare} LPS at BC1 exit for {\sc impact-z}'s (a-d) and {\sc csrtrack}'s (e-h) 3D models, for 3.2-nC (a,e), 1-nC  (b,f), 250-pC (c,g), and 20-pC (d,h) bunch charges, zoomed in to show details, for ${\cal C}= 5.2$~m$^{-1}$.  (Red line) longitudinal current projection, with arbitrary scale and offset.  Note that the horizontal and vertical axis ranges are different for each plot. The ordinates $z>0$ correspond to the head of the bunch. (color online)}
\end{figure}

\begin{figure}[h!]
\centering
\includegraphics[width=2.92087in]{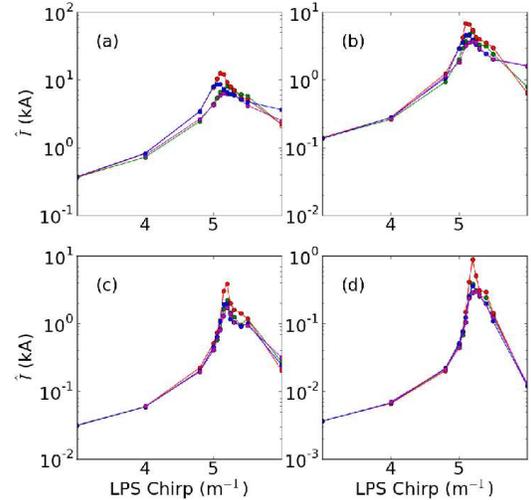}
\caption{\label{fig:currents} Peak currents $\hat{I}$ versus energy chirp for {\sc impact-z}'s SC+CSR (green), {\sc impact-z}'s SC (blue), {\sc csrtrack}'s 1DP (red), and {\sc csrtrack}'s P2P (magenta) models, for 3.2-nC (a), 1.0-nC (b), 250-pC (c), and 20-pC (d) bunch charges. (color online)}
\end{figure}

The LPS 1.0-m downstream of the B4 dipole simulated with {\sc impact-Z} (SC + one-dimensional CSR models) and {\sc csrtrack} (P2P model) are  shown in Fig.~\ref{fig:LPScompare} [(a)-(d)] and [(e)-(f)] respectively for the four charges listed in Tab.~\ref{tab:ChargeParameters}. Despite the vastly different algorithms used by these two programs, the LPS distributions displayed very similar distortions including those at the small-scale levels.  Figure~\ref{fig:currents} summarizes the evolution of peak current  as a function of the initial LPS chirp for the four numerical models. Likewise the longitudinal emittances computed with {\sc impact-z} (SC+CSR) and {\sc csrtrack} (P2P) for the case of maximum compression are in decent agreement; see Tab.~\ref{tab:FinalENZ}.  

\begin{table}[h!]
\caption{\label{tab:FinalENZ} Final normalized longitudinal $\varepsilon_{z}$ at maximum compression (${\cal C}=5.2$~m$^{-1}$) simulated with {impact-z} (SC+CSR) and {\sc csrtrack} (P2P).}
\begin{center}
\begin{tabular}{c c c}
\hline 
  & {\sc impact-z}   & {\sc csrtrack}  \\
  & SC+CSR   & P2P  \\
Q (nC) &  $\varepsilon_{z}$ ($\mu$m)  &  $\varepsilon_{z}$ ($\mu$m) \\
\hline \hline 
3.2 & 267  & 261   \\
1.0 & 118  & 105   \\
0.250 &  61.5  & 57.8  \\
0.020 &  10.5   & 11.6  \\
\hline 
\end{tabular}
\end{center}
\end{table}

\begin{figure}[h!]
\centering
\includegraphics[width=0.48\textwidth]{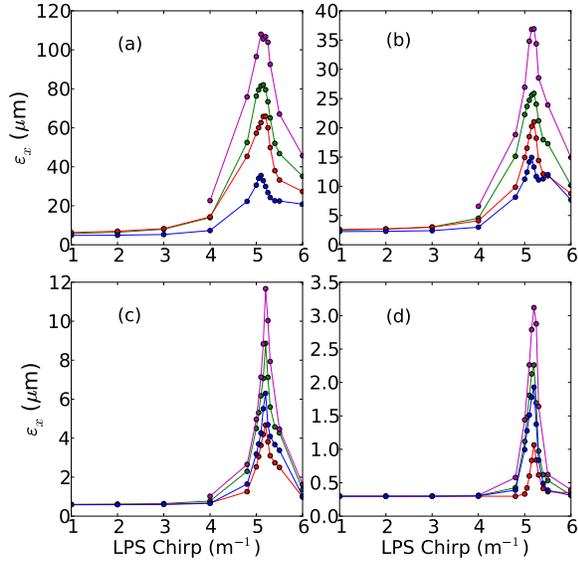}
\caption{\label{fig:ENX} Final horizontal emittances for each of the different bunch charges with {\sc impact-z}'s SC-only model (blue), {\sc csrtrack}'s 1D CSR-model (Red), {\sc impact-z}'s SC+CSR model (Green), and {\sc csrtrack}'s P2P model (magenta), for 3.2-nC (a), 1-nC  (b), 250-pC (c), and 20-pC (d) bunch charges.  (color online)}
\end{figure}

\begin{figure}[h!]
\centering
\includegraphics[width=0.48\textwidth]{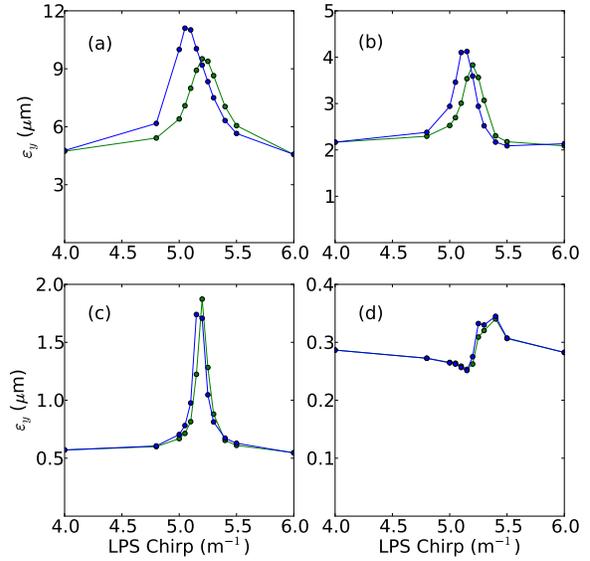}
\caption{\label{fig:ENY} Final vertical emittances for each of the different bunch charges with {\sc impact-z}'s SC-only model (blue) and {\sc impact-z}'s SC+CSR model (Green), for 3.2-nC (a), 1-nC  (b), 250-pC (c), and 20-pC (d) bunch charges.  {\sc Csrtrack} does not compute vertical forces, so the emittance remains roughly constant along the bunch compressor. (color online)}
\end{figure}

The transverse-emittance after compression is shown in Fig.~\ref{fig:ENX}.  {\sc Csrtrack}'s P2P model consistently offers the greatest emittance growth, followed by {\sc impact-z}'s SC+CSR model, as these are the only two models that account for both SC and CSR effects.  However, the P2P model also includes transverse CSR forces and a more elaborate model for longitudinal CSR.  Our previous study~\cite{ProkopBC1TM} showed that the influences on final emittance from using too-few macroparticles as well from as the randomization in the down-sampling of initial distributions were both much smaller than the discrepancy between the SC+CSR and P2P models.  The emittance growth observed from {\sc csrtrack}'s 1D and {\sc impact-z}'s SC-only model, indicates that CSR accounts for most of the emittance dilution at higher charge. For the low-charge simulations ($Q=250$ and 20~pC) the relative importance is reversed, with SC contributing more to the emittance degradation than CSR.  Of the models presented here, only {\sc impact-z}'s include SC in both the vertical and the horizontal planes, and are shown in Fig.~\ref{fig:ENY}.  As vertical emittance growth is entirely the result of SC, the inclusion of {\sc impact-z}'s CSR model reduces vertical emittance growth due to the reduced compression.

\begin{figure}[h!]
\centering
\includegraphics[width=0.48\textwidth]{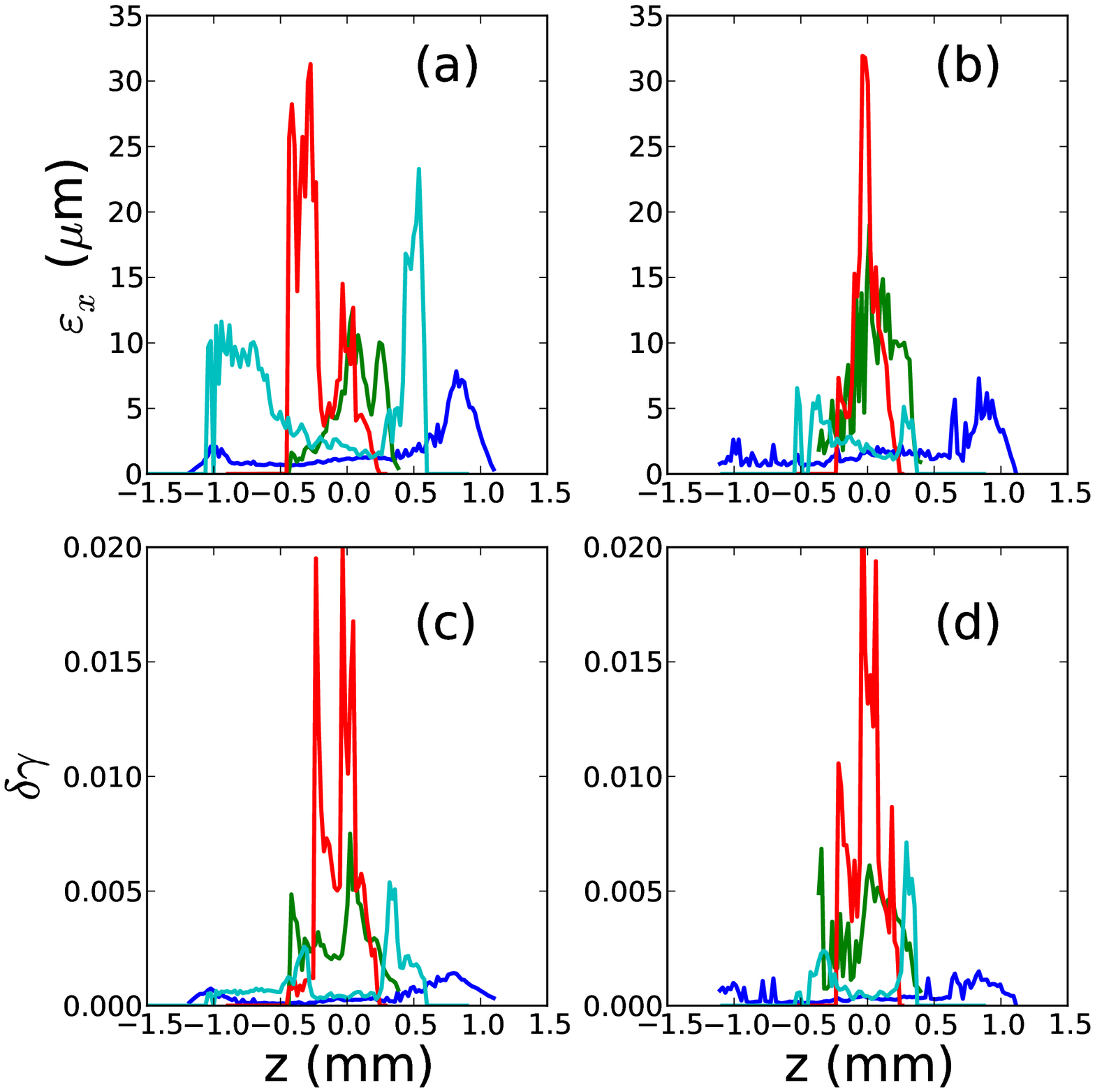}
\caption{\label{fig:sliceparam} Example of final normalized transverse slice emittances (top row) and energy spread (bottom row)  evolution within a 1-nC bunch for four cases of compression ${\cal C}=4.0$, 5.0, 5.25, and 6.0~m$^{-1}$ respectively shown as blue, green, red and turquoise traces). Plots (a) and (c) correspond to {\sc impact-z} simulations while plots (b) and (d) are results from  {\sc csrtrack}'s P2P model.  Emittance and energy spread values associated to slices that contain too-few number of macroparticle for meaningful statistical analysis are set to zero.  The heads and tails of the bunches are sparsely populated (see Fig.~\ref{fig:LPScompare} for reference), particularly for the P2P simulations which use only 5~\% of the number of particles used in the {\sc impact-z} simulations. }
\end{figure}

\begin{figure}[h!]
\centering
\includegraphics[width=0.48\textwidth]{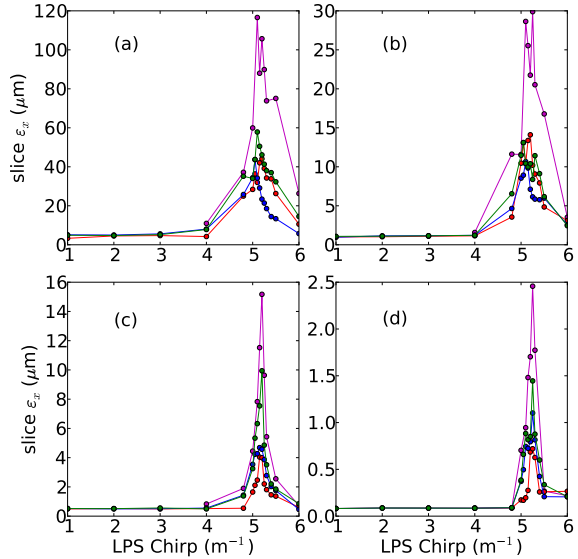}
\caption{\label{fig:SliceENX} Final normalized transverse slice emittances $\varepsilon_{x,M}$ in the slice with the highest peak current computed  with {\sc impact-z}'s SC-only model (blue), {\sc csrtrack}'s 1D CSR-model (red), and {\sc impact-z}'s SC+CSR model (green), and {\sc csrtrack}'s P2P model (magenta), for 3.2-nC (a), 1-nC  (b), 250-pC (c), and 20-pC (d) bunch charges.}
\end{figure}

Finally, the evolution of the  slice parameters during compression were explored. For this analysis the beam is divided into axial slices of equal longitudinal length $\delta z=20$~$\mu$m. A statistical analysis on the population contributing to each slice was performed to yield the slice emittances, energy spread and peak current. A comparison of the slice bending-plane emittance and energy spread between {\sc impact-z} and {\sc csrtrack}'s P2P model of slice-emittance as the chirp is varied appears in Fig.~\ref{fig:sliceparam}. The level of agreement between the two programs is of the same order as what observed for the bunch parameters.  Fig.~\ref{fig:SliceENX} summarizes the evolution of the slice horizontal emittance in the slice with the highest peak current within the bunch.  When the beam is greatly over-compressed (${\cal C}= 6.0$~m$^{-1}$), the LPS may be double-peaked, and the transverse brightness as defined here may not be an appropriate measure of the bunch's utility. The curve demonstrates how little slice emittance growth occurs for partial compression.

\section{Expected beam dilution and trade-offs}
\begin{figure}[h!]
\centering
\includegraphics[width=2.92087in]{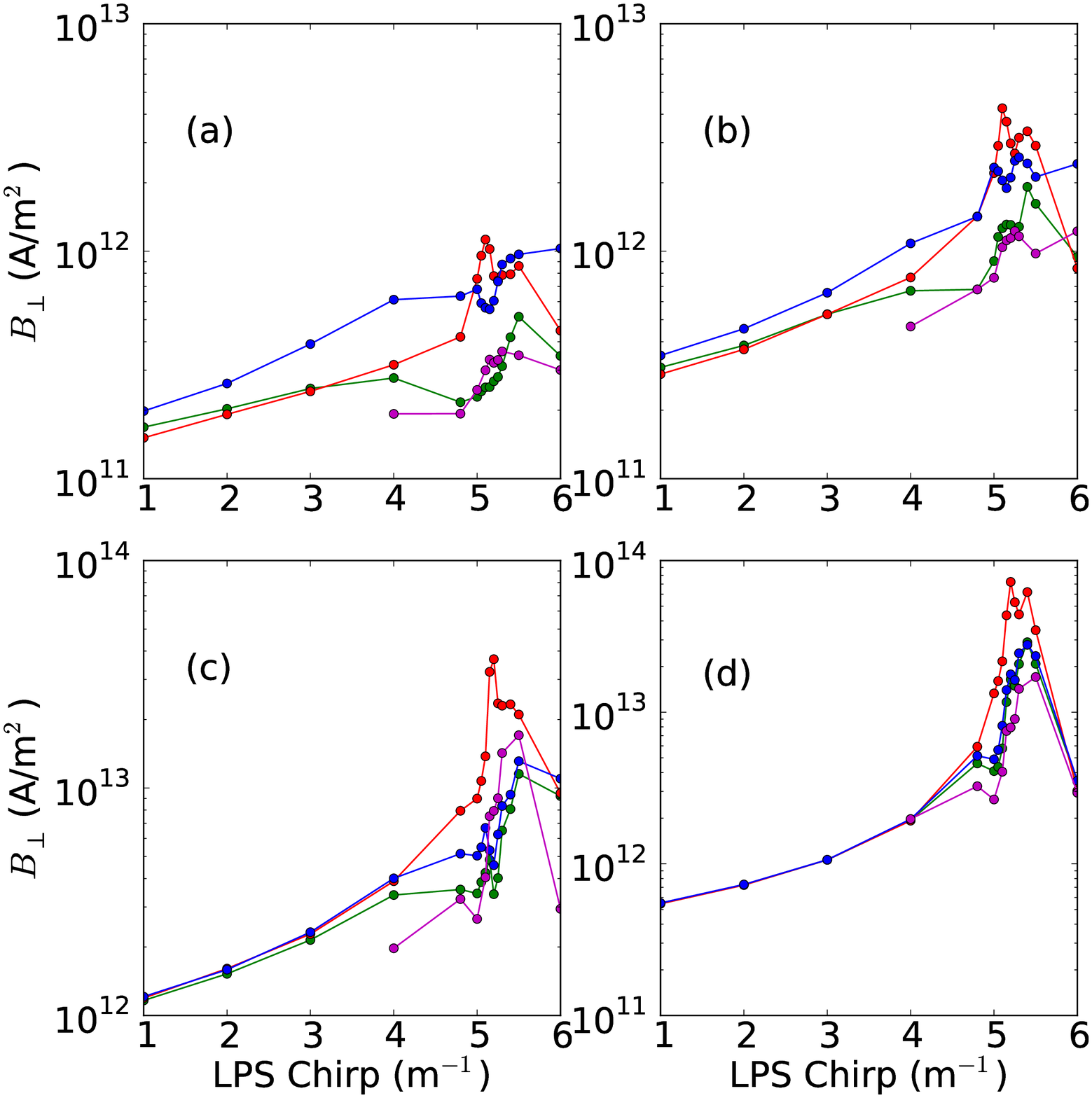}
\caption{\label{fig:peakbrightness} Peak transverse brightness $B_{\perp}$=$\frac{\hat{I}}{4 \pi^2 \varepsilon_{x} \varepsilon_{y}}$ versus energy chirp for {\sc impact-z}'s SC+CSR (green), {\sc impact-z}'s SC (blue), {\sc csrtrack}'s 1DP (red), and {\sc csrtrack}'s P2P (magenta) models, for 3.2-nC (a), 1.0-nC (b), 250-pC (c), and 20-pC (d) bunch charges. (color online)}
\end{figure}

A large number of accelerator applications require beams with high-peak-currents and low-transverse-emittances. These requirements conflict with each other as collective effects, which dilute the beam's phase space and emittances, increase with peak current.  A commonly-used figure of merit is peak transverse brightness $B_{\perp} \equiv \frac{\hat{I}}{4 \pi^2 \varepsilon_{x} \varepsilon_{y}}$~\cite{BJC}.  Figure~\ref{fig:peakbrightness} summarized the evolution of $B_{\perp}$ as function of the LPS chirp for the four cases of bunch charges. The figure combines the data provided in Fig.~\ref{fig:currents} and Fig.~\ref{fig:ENX}. Despite the lower-charge bunch result in smaller peak current at lower charges (see Fig.~\ref{fig:currents}), the transverse brightness increases with lower bunch charges. The main factor at play in this reduction is the lower initial transverse emittances, and more importantly, the reduced dilution of the transverse emittances during compression in BC1 due to the weaker collective effects (CSR and SC).

\begin{figure}[h!]
\centering
\includegraphics[width=2.92087in]{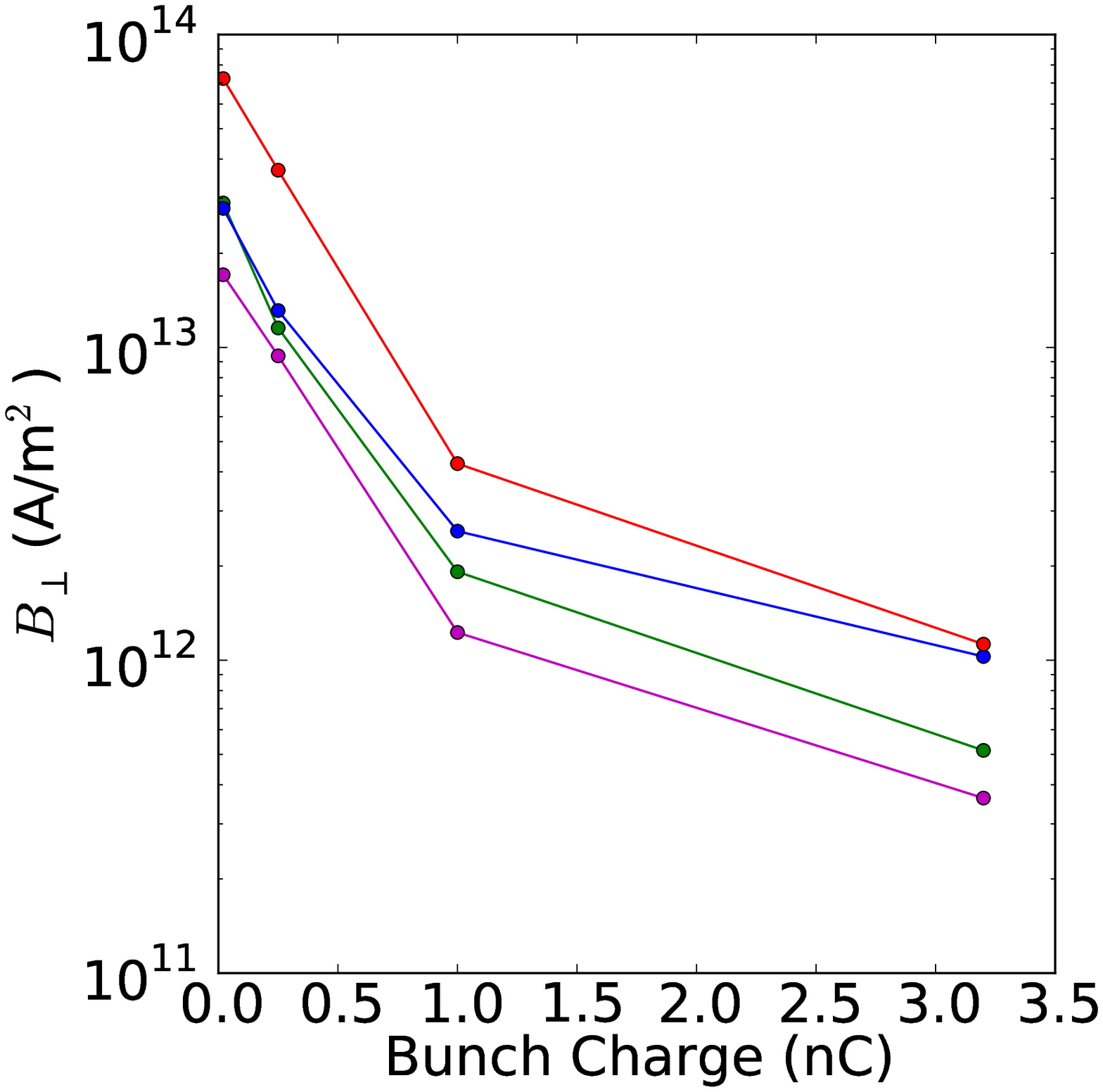}
\caption{\label{fig:MaxBrightness} Maximum peak transverse brightness $B_{\perp}$=$\frac{\hat{I}}{4 \pi^2 \varepsilon_{x} \varepsilon_{y}}$ versus bunch charge for {\sc impact-z}'s SC+CSR (green), {\sc impact-z}'s SC (blue), {\sc csrtrack}'s 1DP (red), and {\sc csrtrack}'s P2P (magenta) models. Each data point is a maximum from each line in Fig.~\ref{fig:peakbrightness}. (color online)}
\end{figure}

In addition we note that the maximum achieved transverse brightness does not necessarily occur at maximum compression.  This is due to the larger peak currents at maximum compression that drive collective effects, wherein the relative emittance growth driven by collective effects is greater than the relative increase in peak current, a trade-off similar to that of going down to lower bunch charges. A summary of the achieved maximum value of $B_{\perp}$ appears in Fig.~\ref{fig:MaxBrightness}.  

\begin{figure}[h!]
\centering
\includegraphics[width=2.92087in]{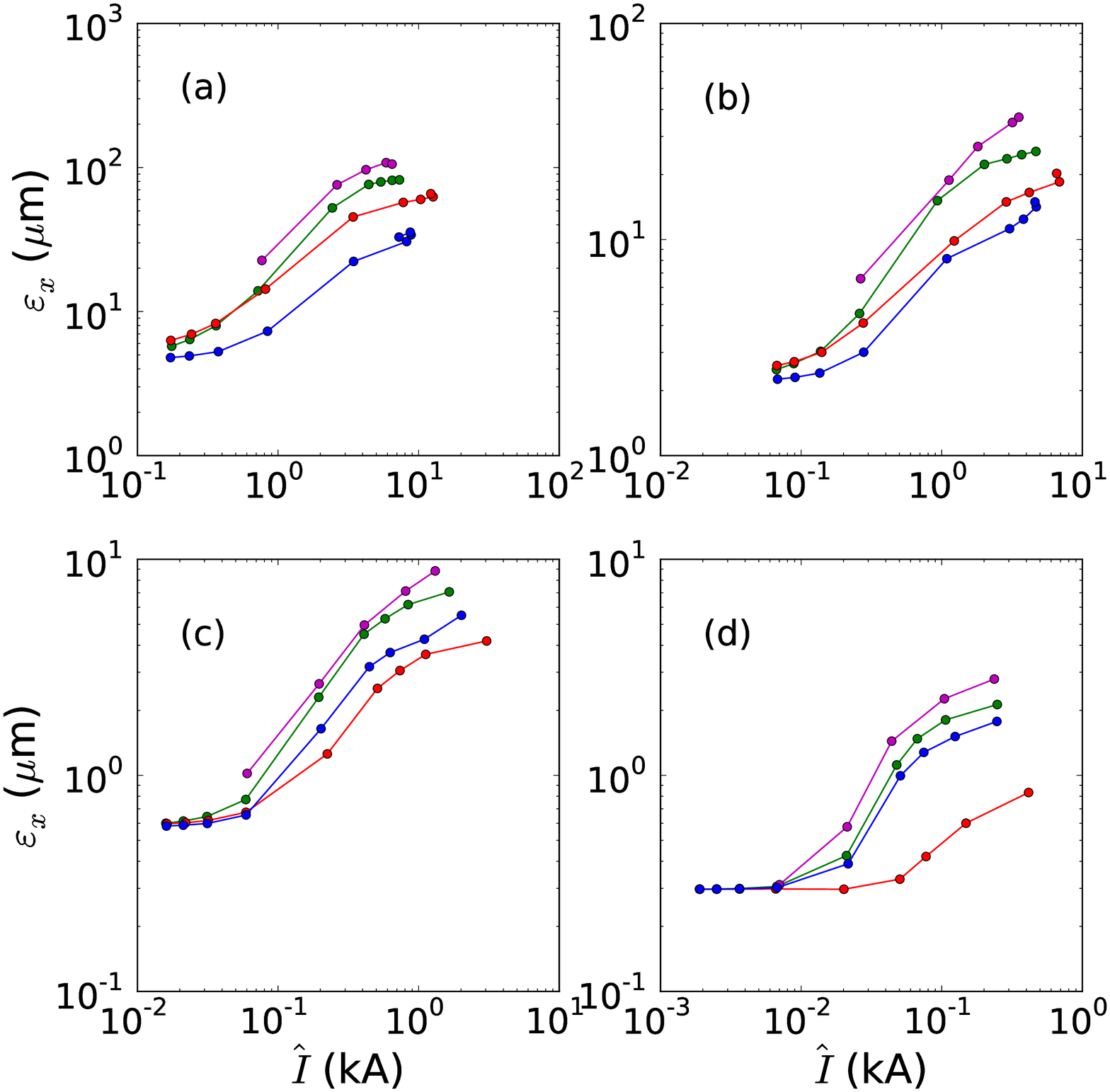}
\caption{\label{fig:ENXpeakTradeOff} Final normalized transverse emittances $\varepsilon_{x}$ versus peak currents $\hat{I}$ for LPS chirps using {\sc impact-z}'s SC+CSR (green), {\sc impact-z}'s SC (blue), {\sc csrtrack}'s 1DP (red), and {\sc csrtrack}'s P2P (magenta) models, for 3.2-nC (a), 1.0-nC (b), 250-pC (c), and 20-pC (d) bunch charges.  Only data corresponding to chirp values ${\cal C} \in[1.0,5.2]$~m$^{-1}$ are displayed. (color online)}
\end{figure}

The trade-off between obtained peak current and $\varepsilon_{x}$ is shown in Fig.~\ref{fig:ENXpeakTradeOff}; only data associated to LPS chirp up to maximum compression at ${\cal C}\le 5.2$~m$^{-1}$ are displayed, as over-compression results in lower peak currents with generally larger emittance dilutions.

The simulation data points to several conclusions about the parametric trade-offs that must be considered.  First, the emittance growth, particularly the slice emittance, is greatly reduced at lower degrees of compression, particularly ${\cal C}<4$~m$^{-1}$, which corresponds to compression to around one-third of the initial bunch length.  Second, using lower bunch-charges is preferred for experiments that require high transverse brightnesses, due to the lower emittance growth from collective effects justifying the lowered peak current.  For the 20-pC bunch charge, the regime with ${\cal C} < 4.0$~m$^{-1}$ results in horizontal emittance growth that is under 10\% of the initial horizontal emittance, regardless of which of the simulation codes is used.  Under-compression is discussed further in section 6.1.

\section{Applications}
The main motivation that led to the inclusion of BC1 in the ASTA's photoinjector is to provide a weak compression necessary to avoid significant energy spread to be accumulated during acceleration in subsequent cryomodule. It is anticipated that a second-stage bunch compressor will eventually be installed at higher energy to compress the bunch to high peak current while mitigating phase-space dilution. However, in light of the studies presented in the previous Sections it is worth investigating other possible applications of the BC1 compressor as discussed below. 

\subsection{Multi-stage bunch compression}
One of the motivations for exploring low-energy bunch compression is to moderately compress the bunches before injection in an accelerating structure and then further compress at high energy~\cite{BCTTF1}. In Ref.~\cite{BCTTF1}, the low-energy bunch compression was accomplished at $\sim15$~MeV and resulted in intolerable beam degradation at the Tesla Test Facility I (TTF1) and consequently dismantled from the beamline during its upgrade as the Free electron LASer in Hamburg (FLASH) user facility. The requirement on the first-stage bunch compression is to provide bunch lengths that satisfy  $\sigma_z\ll \lambda/(2\pi)$ (where $\lambda$ is the wavelength of the RF wave associated with the subsequent accelerating structure). The latter condition insures that no significant LPS quadratic distortion is imparted during acceleration in the subsequent linac. For the L-band linac of ASTA this sets the upper requirement $\sigma_z\le 800$~$\mu$m~\cite{BCTTF1}. For a Gaussian distribution this requirement corresponds to a $\hat{I}\simeq 500$~A at $Q=3.2$~nC and ${\cal C} \sim 3.7$~m$^{-1}$ resulting in the tolerable bending-plane emittance dilution as shown in Tab.~\ref{tab:PartialCompression} along with compression to $\sigma_z\simeq 800$~$\mu$m for each of the other bunch charges.  Relative emittance growth is significantly smaller for 1-nC bunch charge and below.


\scriptsize
\begin{table}
\caption{\label{tab:PartialCompression} Final normalized transverse $\varepsilon_{x/y}$ and longitudinal $\varepsilon_{z}$ emittances for the four cases of charge considered in this paper. The initial LPS chirp (${\cal C}$) was optimized for each charge to yield a final bunch length $\sigma_z\simeq 800$~$\mu$m, based on the scan presented in Fig.~\ref{fig:IZ_sigz}. The simulations were performed with {\sc impact-z}'s SC+CSR model. The values displayed in this Table should be compared with the pre-compression values summarized in Tab.~\ref{tab:ChargeParameters}.}
\begin{center}
\begin{tabular}{c c c c c }
\hline 
Q (nC) & ${\cal C}$ (m$^{-1})$ &  $\varepsilon_{x}$ ($\mu$m) & $\varepsilon_{y}$ ($\mu$m) &  $\varepsilon_{z}$ ($\mu$m)    \\
\hline \hline 
3.2 & 3.7 & 10.56 & 4.24 & 88.7   \\
1.0 & 3.2 & 3.03 & 1.98 & 34.2   \\
0.250 & 3.1 & 0.623 & 0.594 & 14.7  \\
0.020 & 1.9 & 0.296 & 0.293 & 1.83  \\
\hline 
\end{tabular}
\end{center}
\end{table}
\normalsize

\subsection{High peak current production }

Despite their relatively poor transverse emittance, fully compressed bunches could be used to generate copious amounts of radiation via a given electromagnetic process. The spectral-angular fluence emitted by a bunch of $N\gg 1$ electrons from  any electromagnetic process  is related to the single-electron spectral fluence,  $\frac{d^2W}{d\omega d\Omega}\big|_1$,  via 
\begin{eqnarray}
\frac{d^2W}{d\omega d\Omega }\big|_N\simeq \frac{d^2W}{d\Omega d\omega}\big|_1 [N+N^2|S(\omega)|^2],
\end{eqnarray}
 where $\omega\equiv 2\pi f$ ($f$ is the frequency) and $S(\omega)$, the bunch form factor (BFF), is the intensity-normalized Fourier transform of the normalized charge distribution $S(t)$~\cite{saxon}. The former equation assumes the bunch can be approximated as a line charge distribution and is practically valid as long as the rms bunch duration $\sigma_t$ and transverse size $\sigma_{\perp}$ satisfy $\sigma_{\perp} \ll c \sigma_t/\gamma$ where $\gamma$ is the Lorentz factor and $c$ is the velocity of light. When the BFF approaches unity, $\frac{d^2W}{d\omega d\Omega }\big|_N \propto N^2$ and the radiation is termed ``coherent radiation".  
 
\begin{figure}[h!]
\centering
\includegraphics[width=0.48\textwidth]{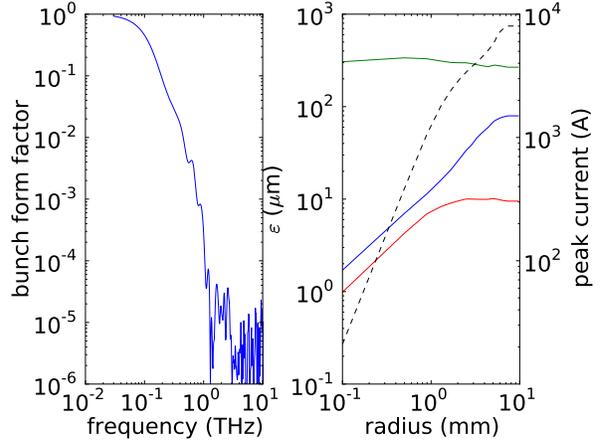}
\caption{\label{fig:radius} Bunch form factor associated to the 3.2-nC fully-compressed electron bunch (left plot) and final horizontal (blue), vertical (red), and longitudinal (green) normalized emittances (left vertical axis) and peak current (dashed black line, right axis) of the bunch within a selected transverse radius (right plot). These simulations were performed near maximum compression with ${\cal C}=5.2$~m$^{-1}$ and 3.2-nC. (color online)}
\end{figure}

 At ASTA the availability of a superconducing linac coupled with a non-interceptive radiation-generation mechanism (e.g. diffraction radiation~\cite{dr}) could lead to the production of single-cycle THz pulses repeated at 3 MHz over 1-ms. As an example we consider the worst-case scenario of a fully compressed 3.2-nC bunch; the dependency of the BFF over frequency appears in Fig.~\ref{fig:radius} (left plot). The BFF starts to take off at frequency lower than $f \simeq 1$~THz thereby supporting the generation of coherently-enhanced radiation at these frequencies. A limitation might come from the large transverse emittance that would prevent the beam to be focused to a transverse spot RMS size below the required $\sigma_{\perp}/\gamma$ value. However a statistical analysis indicates that the central part of the beam containing approximately 15\% of the beam population (or 500-pC out of the original 3.2-nC bunch) has emittances below 10~$\mu$m resulting in beam $\sigma_{\perp}\le 100$~$\mu$m; see Fig.~\ref{fig:radius} (right plot).  It should be pointed out that the lower-charge cases investigated in the previous section would result in shorter pulses with associated BFFs that contain higher-frequency content (see also Fig.~\ref{fig:IZ_sigz}).

\subsection{Compressed flat-beam generation}
An important asset of the ASTA photoinjector is its capability to generate beams with high-transverse emittance ratios known as flat beams. Immersing the photocathode in a magnetic field introduces a canonical angular momentum $\meanavg{L}=eB_0\sigma_c^2$, with $B_0$ the magnetic field on the photocathode surface, and $\sigma_c$ the RMS transverse size of the drive-laser spot on the photocathode~\cite{FlatBeamSource}. As the beam exits the solenoidal field provided by lenses L1 and L2, the angular momentum is purely kinetic resulting in a beam coupled in the two transverse planes. Three skew quadrupoles in the beamline can apply the torque necessary to cancel the angular momentum~\cite{FlatBeamSource2,FB3}. As a result, the final beam's transverse emittance partition is given by
\begin{equation}
(\varepsilon_{x,i}, \varepsilon_{y,i})= \left(\frac{\varepsilon_u^2}{2 \beta \gamma
 \mathcal{L}}, 2 \beta \gamma \mathcal{L}\right)  \label{eq:kjk}
,
\end{equation}
where $\varepsilon_u$ is the normalized uncorrelated emittance of the  magnetized beam prior to the
transformer, $\beta$ and $\gamma$ the Lorentz factors, ${\cal L}\equiv \meanavg{L}/2p_z$, and $p_z$ is the longitudinal momentum. Note
that the product $\varepsilon_{x,i} \varepsilon_{y,i}  =(\varepsilon_u)^2$. If compressed these flat beams may have applications in Smith-Purcell FELs~\cite{spfel} or for beam-driven acceleration techniques using asymmetric structures~\cite{daniel}. It may also  be possible to mitigate the emittance growth in BC1 by having a beam that is wide in the direction of the chicane bend. 

\begin{figure}[ht!]
\centering
\includegraphics[width=0.49\textwidth]{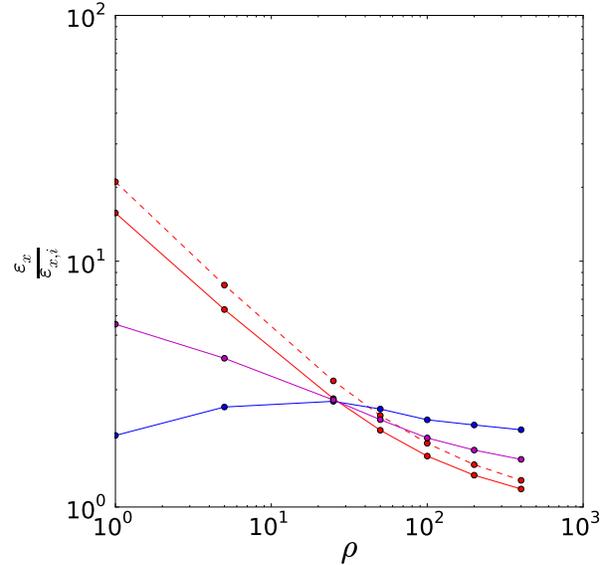}
\caption{Bending plane transverse emittance $\frac{\varepsilon_{x}}{\varepsilon_{x,i}}$ growth in BC1 (red) simulated with {\sc csrtrack} (dashed line) and {\sc impact-z} (solid lines) as functions of the initial emittance ratio $\rho\equiv\frac{\varepsilon_{x,i}}{\varepsilon_{y,i}}$. Corresponding {\sc impact-z} results for the vertical emittance ($\frac{\varepsilon_{y}}{\varepsilon_{y,i}}$, blue solid line), and four-dimensional transverse emittance ($\frac{\varepsilon_{4}}{\varepsilon_{4,i}}$, magenta solid line).  (color online)}\label{fig:FlatScan}
\end{figure}

 In this section, we explore the behavior of flat beams in the low-energy bunch compressor at ASTA, for  the different initial emittance ratios $\rho\equiv \varepsilon_{x,i}/\varepsilon_{y,i}$.  In order to produce these bunches, we took the 3.2-nC bunch presented earlier and numerically scaled the macroparticle coordinates to produce the desired  transverse emittance ratios while constraining the product  $\varepsilon_{x,i} \varepsilon_{y,i}  =5^2$~$\mu$m$^2$.  Due to the large transverse aspect ratio of the bunches, the criterion given in Eq.\ref{eq:DerbenevCriterion} is generally not satisfied and it is therefore anticipated that the projected CSR model is inadequate, thus we use {\sc CSRtrack}'s P2P model to simulate the flat beams and neglect the 1DP model.  The parameters used for flat beam simulations follow those used in the previous section, with the exception of the macroparticle horizontal size used in the {\sc csrtrack} P2P model. Due to the much greater transverse dimension we set $\sigma_{h}=0.2$~mm. In addition, {\sc impact-z} SC+CSR simulations were also performed to evaluate the emittance growth in the vertical plane.  The simulated emittance growth is shown in Fig.~\ref{fig:FlatScan} for a 3.2-nC bunch with an initial LPS chirp of  ${\cal C}= 5.2$~m$^{-1}$. As expected the relative emittance dilution is reduced as the initial emittance ratio $\rho$ increases. The agreement between {\sc csrtrack} and {\sc impact-z} for the bending-plane emittance dilution is remarkable (within $\sim 30$~\%) given the large transverse horizontal beam sizes. {\sc Impact-z} predicts that the vertical emittance  increases by a factor of 1.5 to 1.8 over the range of considered initial emittance ratios $\rho \in[1, 500]$. The four-dimensional transverse emittance growth $\varepsilon_4\equiv \sqrt{\epsilon_x\epsilon_y}$ is mitigated for the larger initial flat-beam emittance ratios. 


\subsection{Double-bunch generation}
\begin{figure}[h]
\centering
\includegraphics[width=0.48\textwidth]{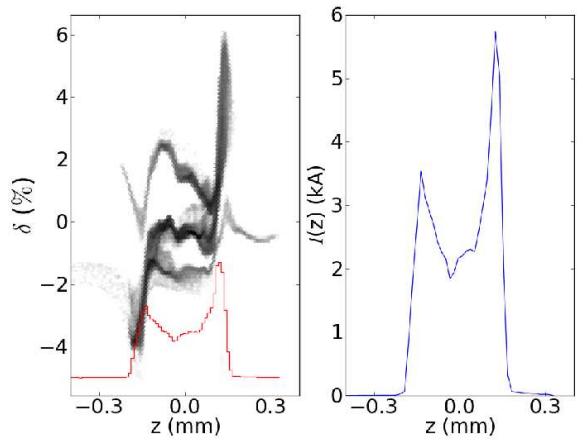}
\caption{\label{fig:doublebunch} LPS distribution (gray colormap in left plot) and current projection (red trace and right plot) associated to over-compressed bunches with an incoming LPS chirp of ${\cal C}=5.5$~m$^{-1}$. The ordinates $z>0$ correspond to the head of the bunch. (color online)}
\end{figure}

The production of shaped electron bunches has a large number of applications including the investigation of wakefield and beam-driven acceleration techniques. Operating the low energy bunch compressor with LPS chirp ${\cal C}>5.5$~m$^{-1}$ leads to over compression and results in a structured longitudinal charge distribution. Figure~\ref{fig:doublebunch} confirms, for the case of $Q=3.2$~nC, that a bi-modal distribution could be generated with a separation between its peaks ($\sim 300$~$\mu$m) consistent with requirements from beam-driven acceleration such as plasma-wakefield and dielectric-wakefield acceleration techniques. In addition the distance between the peaks could be controlled to some degree by slight changes over the initial LPS chirp. The full-bunch and slice-at-peak-current horizontal emittance at ${\cal C} \sim 5.5$~m$^{-1}$ are 67.0 and 75.1~$\mu$m, respectively, compared to 106 and 107~$\mu$m for the maximum-compression case (${\cal C} \sim 5.2$~m$^{-1}$). These bending-plane normalized emittances of $\sim 75$~$\mu$m can still be focused to a sub-mm or sub-100-$\mu$m transverse spot size at respectively $\sim 40$~MeV and $\sim 250$~MeV (the latter energy corresponds to acceleration of the 40 MeV beam into one of ASTA accelerating cryomodules).

\section{Summary}

In this paper we presented numerical studies of a low-energy magnetic bunch compressor similar to the one being installed at the ASTA facility at Fermilab. Our results indicate that low-energy compression can be a viable path as a first stage compressor for a multistage compression scheme. In addition the capability of such a low energy compressor to provide high-peak currents at low energy ($\sim 40$~MeV) could have important applications when combined with the long-macropulse capability of the ASTA's superconducting accelerating module such as, e.g. the production of single-cycle THz radiation from diffraction radiation. As part of our investigation we used several computer programs and observed an acceptable agreement when simulating the evolution of the LPS distributions during compression. The simulated transverse-phase-space parameters downstream of the bunch compressor have discrepancies that are inherent to the capabilities of each of the model but provide similar emittance growth and trade-off curves. Based on these observations, simple (and faster) models could be used to optimize the bunch compression design with a quick turn around while first-principle model could provide high-fidelity simulations of the optimized designs. The installation and commissioning of the BC1 bunch compressor along with the available diagnostics at ASTA will provide a unique experimental platform for benchmarking the simulation codes currently available in regimes where the CSR and SC effects can play similar roles.

\section*{Acknowledgments}
This work was supported by LANL Laboratory Directed Research and Development (LDRD) program, project 20110067DR and by the U.S. Department of Energy under Contract No. DE-FG02-08ER41532 with Northern Illinois University and Contract No. DE-AC02-07CH11359 the Fermi Research Alliance, LLC.

\end{document}